# Evaluation of peak wall stress in an ascending thoracic aortic aneurysm using FSI simulations: effects of aortic stiffness and peripheral resistance

Rossella Campobasso[a], Francesca Condemi[a*], Magalie Viallon[b,c], Pierre Croisille[b,c], Salvatore Campisi[a,b], Stéphane Avril[a]

a. Mines Saint-Etienne, Univ Lyon, Univ Jean Monnet, INSERM, U 1059 Sainbiose, Centre CIS, F - 42023 Saint-Etienne France
b. Centre Hospitalo-Universitaire, Saint-Etienne, France.
c. Univ Lyon, UJM-Saint-Etienne, INSA, CNRS UMR 5520, INSERM U1206, CREATIS, F-42023, Saint-Etienne, France
* now at Department of Electrical and Computer Engineering, University of Toronto, Toronto, Canada

**Abstract.**
**Purpose.** It has been reported clinically that rupture or dissections in thoracic aortic aneurysms (TAA) often occur due to hypertension which may be modelled with sudden increase of peripheral resistance, inducing acute changes of blood volumes in the aorta. There is clinical evidence that more compliant aneurysms are less prone to rupture as they can sustain such changes of volume. The aim of the current paper is to verify this paradigm by evaluating computationally the role played by the variation of peripheral resistance and the impact of aortic stiffness onto peak wall stress in ascending TAA.
**Methods.** Fluid-Structure Interaction (FSI) analyses were performed using patient-specific geometries and boundary conditions derived from 4D MRI datasets acquired on a patient. Blood was assumed incompressible and was treated as a non-Newtonian fluid using the Carreau model while the wall mechanical properties were obtained from the bulge inflation tests carried out in vitro after surgical repair. The Navier Stokes equations were solved in ANSYS Fluent. The Arbitrary Lagrangian Eulerian formulation was used to account for the wall deformations. At the interface between the solid domain and the fluid domain, the fluid pressure was transferred to the wall and the displacement of the wall was transferred to the fluid. The two systems were connected by the System Coupling component which controls the solver execution of fluid and solid simulations in ANSYS. Fluid and solid domains were solved sequentially starting from the fluid simulations.
**Results.** Distributions of blood flow, wall shear stress and wall stress were evaluated in the ascending thoracic aorta using the FSI analyses. We always observed a significant flow eccentricity in the simulations, in very good agreement with velocity profiles measured using 4D MRI. The results also showed significant increase of peak wall stress due to the increase of peripheral resistance and aortic stiffness. In the worst case scenario, the largest peripheral resistance ($10^{10}$ kg.s.m$^{-4}$) and stiffness (10 MPa) resulted in a maximal principal stress equal to 702 kPa, whereas it was only 77 kPa in normal conditions.
**Conclusions.** This is the first time that the risk of rupture of an aTAA is quantified in case of the combined effects of hypertension and aortic stiffness increase. Our findings suggest that a stiffer TAA may have the most altered distribution of wall stress and an acute change of peripheral vascular resistance could significantly increase the risk of rupture for a stiffer aneurysm.

1. **Introduction**

Ascending thoracic aortic aneurysms (aTAA) currently represent the 19th cause of deaths in the world [1]. An aTAA is an abnormal dilatation of the aortic wall which grows most of the time in a silent manner and which can end up into a catastrophic rupture. To prevent aTAA rupture, prophylactic surgery is recommended, whereby the risk of mortality can be as great as 5% [2]. The gold standard for deciding a surgical intervention is based on the "maximum diameter criterion", which is the maximum orthogonal diameter of the vessel with a critical threshold of 5.5 cm [3]. However, for aneurysms with a diameter smaller than 5.5 cm, negative outcomes (rupture, dissection and death) before surgical repair do exist, with an incidence of 5-10% [1]. There is therefore a pressing need to improve the diagnosis tools and to identify patient-specific guidelines for planning surgical repair [4].

It is now widely acknowledged that the ascending thoracic aorta is characterized by a unique bio-chemo-mechanical environment that may play a role in its susceptibility to aTAA and the risk of dissection and rupture. As the main sensor of this bio-chemo-mechanical environment, vascular smooth muscle cells (vSMCs) play a crucial role in the pathogenesis of aTAA, as recently reported in several review articles [5], [6]. Three aetiologies predominate in human aTAA: (i) genetic causes in heritable familial forms [7], (ii) an association with bicuspid aortic valves, and (iii) a sporadic degenerative form linked to the aortic aging process [8]. Whatever the aetiologies, aTAAs are characterized by elastin degradation (proteolytic injury), loss of vSMCs, accumulation of highly hydrophilic glycosaminoglycans (GAGs) and an increase in wall permeability leading to transmural advection of plasma proteins which could interact with vSMCs and components of the extracellular matrix (ECM). The locally disturbed aortic hemodynamics is thought to be related to these effects [8-12]. Most of previously cited studies have focused on characterizing the role of individual factors, i.e. stiffness, wall shear stress (WSS), vSMCs phenotype, or individual gene mutations [13]. However, even if these individual factors offer a clear picture of aTAA natural history, it is clear that individual factors cannot predict failure risk at a patient-specific level.

Patient-specific rupture risk prediction requires determining when the stress applied to the aortic wall locally exceeds its strength. Finite-element analyses can be used to estimate the local distribution of the stress applied by the blood pressure onto the aortic wall [12, 14-17]. An open question is still to estimate the patient-specific strength, which can vary from a few tenths of MPa to a few units of MPa from one individual to another [18-22]. Another open question is that mean physiological wall stresses (the stresses which does not exceed the wall strength) acting on pathologic aortas were found to be far from rupture, with factors of safety (defined as the ratio of tensile strength to the mean wall stress) larger than six [23].

Rupture risk prediction could also be achieved by determining when the stretch applied to the tissue exceeds its extensibility. For instance, Martin et al. [24] defined a new rupture risk criterion (the diameter ratio risk) as the ratio between the current diameter of the aneurysm and the rupture diameter. They showed that the diameter ratio risk increases significantly with the physiological elastic modulus of the artery. This physiological modulus was derived from the Laplace law considering a pressure range of 80–120 mmHg. Our research group [21] proposed a similar rupture risk criterion, namely the stretch ratio risk, defined as the ratio between the physiologic tissue stretch and the maximum stretch (at which the tissue ruptures). To assess the physiologic tissue stretch, we first estimated the average physiologic tissue tension, under in vivo conditions, using the Laplace law. Then the physiologic tissue stretch, corresponding to the average physiologic tissue tension, was deduced from the tension-stretch response of the same aneurysm (collected during the surgical procedures) measured in vitro in a bulge inflation test. We derived the stretch ratio risk criterion for a cohort of 31 patients using this procedure. We also derived the tangent elastic modulus of the aTAA tissues and demonstrated that it is strongly correlated to the stretch ratio risk criterion [21]. This relationship between stiffness and rupture susceptibility could be used clinically to inform about the risk of aTAA rupture as the aortic stiffness can be measured non-invasively in any patient. Given the likely progressive increase in stiffness in response to proteolytic injury, any acute increase in blood pressure could significantly increase wall stress and render aneurysmal vessels more susceptible to failure. Indeed, patients with Marfan syndrome, and similarly for others with aTAA, should avoid strenuous activities that increase blood pressure acutely, such as weight lifting [24, 25], as rupture or dissections in aTAA often occur at a time of severe emotional stress or physical exertion [24].

During hypertension, the cardiac output remains pretty much unchanged while the resistance to blood flow increases leading to elevated blood pressure [26]. To the best of our knowledge, the influence of an acute change of peripheral aortic resistance or the impact of aortic stiffness on the aTAA risk of rupture have never been quantified computationally.

The objective of this paper is to set up an original framework for the fluid structure interaction (FSI) analysis of aTAA patients affected by an acute change of peripheral resistance. The image-guided FSI analysis was developed using patient-specific boundary conditions and was verified against 4D MRI datasets. Then the influence of aortic stiffness and peripheral resistance on aTAA peak wall stress was investigated.

2. **Material and Methods**

**2.1 Data Acquisition and reconstruction of fluid and solid domains**

The study was approved by the Institutional Review Board of the University Hospital center of Saint-Étienne (France). After informed consent, a 59-year-old man presenting a 60 mm diameter aTAA was enrolled. The patient presented a bicuspid aortic valve (BAV) with a moderate aortic valve insufficiency (AI II grade) and a "bovine arch" morphology of the aortic arch [27]. The day before surgical repair, the patient was scanned on a 3T MRI scanner (Siemens Magnetom Prisma) without contrast agent using a 4D flow phase contrast protocol and sequence [28]. The acquisition was performed with a true spatial resolution of 1.9x1.9x2.2 mm$^3$, field of view (FOV) = 360 mm, BW = 740 Hz/pixel, flip angle = 8°, TE/TR/TI = 2.9/39.2/150 ms, venc = 350 cm/s and phase duration = 39.2 ms. A prospective electrocardiogram (ECG) gating was used. The 4D flow MRI data analysis and visualization was performed using cvi$^{42®}$ prototype 4D Flow module (cmr42, Circle Cardiovascular Imaging Inc., Calgary, Canada [29]).

CRIMSON (CardiovasculaR Integrated Modelling and SimulatiON) software was used to reconstruct the fluid domain ($\Omega_f$) from the 4D MRI scan taken at the beginning of the diastolic phase. The fluid domain included the aortic arch, the apico-aortic branches (brachiocephalic artery, BCA, left common carotid artery, LCC, and left subclavian artery, LSUB) and the descending aorta (DescAo).

The thickness of the arterial wall could not be measured *in vivo* as neither CT scans nor MRI have sufficient spatial resolution. For this reason, starting from the boundary of the fluid domain (luminal surface of the wall), the solid domain, denoted $\Omega_s$, was extruded in outer normal direction and by a constant thickness value of 1.5 mm.

**2.2 Numerical simulations**

**2.2.1. Fluid model**

Blood was assumed incompressible and was treated as a non-Newtonian fluid using the Carreau model [30]. The velocity field $\boldsymbol{v_f}$ and the pressure field $p$ across $\Omega_f$ satisfy the transient Navier-Stokes equations which may be written in Arbitrary Lagrangian Eulerian formulation:

$$\begin{cases} \rho_f \frac{\partial \boldsymbol{v_f}}{\partial t} + \rho_f (\boldsymbol{v_f} - \boldsymbol{w}) . \nabla \boldsymbol{v_f} - \nabla . \boldsymbol{\tau_f} + \boldsymbol{\nabla} p = \boldsymbol{0} & (1a) \\ \nabla . \boldsymbol{v_f} = 0 & (1b) \end{cases}$$

where $\rho_f$ is the blood density, $\boldsymbol{\tau_f}$ is the shear stress tensor expressed with respect to strain rates according to the Carreau model, and $\boldsymbol{w}$ is the velocity field of the fluid domain relative to the ALE formulation which satisfies the Laplace equation:

$$\Delta \boldsymbol{w} = 0 \qquad (2)$$

To find an approximate solution to the Navier-Stokes and Laplace equations, the fluid domain was partitioned in tetrahedral elements using Ansys ICEM CFD 17.2 (ANSYS Inc. Canonsburg, PA, USA). As the focus of the study is to evaluate the stress in the aortic walls rather than the WSS, a uniform fine mesh was chosen. Mesh independency was evaluated by testing two meshes: coarse (maximum elements size of 1.5 mm) and fine (maximum elements size of 1 mm). The solution was considered mesh-independent for an error lower than 2% in terms of velocity and pressure. The fine mesh (6.1 M elements, 1.3 M nodes) was used to achieve the results presented in the current work.

Ansys Fluent was used to solve the governing integral equations for the conservation of mass and momentum. Unknown $\boldsymbol{v_f}$, $\boldsymbol{w}$ and $p$ were defined starting from the integration of the governing equations on the individual control volumes and proceeding with linearization and resolution of the resulting system of linear equations, yielding updated values of the unknown variables [31].

After building the mesh, the Navier-Stokes and Laplace equations (Eqs 1a, 1b and 2) were solved in ANSYS Fluent to proceed with the simulations. The flow was assumed laminar.

Boundary conditions were assigned at the boundaries of the fluid domain.

Boundaries corresponding to arterial walls (interface between the fluid and solid domains) were assigned a condition:

$$\boldsymbol{v_f} = \boldsymbol{w} = \frac{\partial \boldsymbol{u_s}}{\partial t} \qquad (3)$$

$\boldsymbol{u_s}$ being the displacement in the solid domain.

Other boundary conditions need to be assigned at the different inlets and outlets of the fluid domain, which are planes defined with a unit normal vector $\boldsymbol{n_{out}}$. Whereas a condition $\boldsymbol{w}.\boldsymbol{n_{out}} = 0$ was assigned for the mesh velocity at these boundaries, conditions deduced from experimental datasets were used for $\boldsymbol{v_f}$.

The patient-specific map of velocity profile was obtained from the 4D flow MRI and was used as inflow boundary condition at the aorta inlet (AAoinlet, Figure 1A).

Previous studies have highlighted the importance of imposing appropriate outlet boundary conditions in hemodynamics numerical simulations. Cheng et al. [32] have used the measured pressure waveform at the outlet of the descending aorta. Pirola et al. [33] have applied outlet boundary conditions by taking into account the interaction between the 3D domain of interest and the remaining part of the vascular system. The strategy of coupling a 3D model to a 0D model addresses three main problems related to the use of 4D MRI flow rate waveforms: 1) the 4D MRI data reduced temporal and spatial resolution, 2) the inflow and outflows phase shifts due to the vessel compliance which is not included in 3D rigid models and 3) the constraint of mass conservation [34]. However, obtaining the realistic flow rate distribution at the outlets by using coupled models requires tuning several parameters which is an expensive and challenging operation. Moreover, the lack of invasive pressure measurements can potentially introduce errors and uncertainties in the results. Therefore, in absence of invasive pressure measurements, we choose to prescribe in vivo hemodynamic quantities as boundary conditions [35, 34]. Finally, at the DescAo, a multi-scale approach was implemented by coupling the 3D domain with a reduced order model. A three-elements Windkessel model was used, which relates the blood flow and blood pressure. Two resistors (impedance $Z_c$ and distal resistance $R$), represented the characteristic resistance of the artery and the peripheral resistance and a capacitor $C$ represented the total systemic arterial compliance [36].

As explicit time integration is unstable for FSI system simulations, the time integration of Navier Stokes equations was performed using a semi-implicit pressure-based solver. The Semi-Implicit Method for Pressure-Linked Equations (SIMPLE) algorithm was used to solve the continuity equations (Eq. 1b) and the linearized momentum equations (deriving from Eq. 1a) in a sequential fashion (instead of being solved simultaneously in a coupled algorithm). A second-order interpolation scheme was chosen for calculating cell-face pressures and, to discretize the convective terms in Eq 1a, a second-order upwind interpolation scheme was applied.

Finally, a second order implicit time advanced scheme was used as transient-time solver and a time step of 0.001 s was chosen for the simulations. The convergence of the solution was assessed for residual errors below $10^{-3}$.

Eq. 2 was solved using the diffusion-based smoothing method. This algorithm moves mesh nodes in response to displacement of boundaries by calculating a mesh velocity using a diffusion equation, i.e. the velocity at the boundary nodes is used as a Dirichlet boundary condition.

### 2.2.2. Solid model

The solid domain $\Omega_s$ was made of the aortic wall and the wall of the 3 branches of the supra aortic trunk. Assuming that wall strains remained infinitesimal throughout a cardiac cycle, the wall constitutive behavior could be linearized and modeled as linear elastic isotropic, satisfying the following constitutive equations:

$$\boldsymbol{\sigma}_s - \boldsymbol{\sigma}_S^0 = \frac{E}{1+v}\boldsymbol{\varepsilon}_s - \frac{Ev}{(1+v)(1-2v)}\text{Tr}(\boldsymbol{\varepsilon}_s)\boldsymbol{I} \qquad (4)$$

where $\boldsymbol{\sigma}_s$ is the stress tensor in the solid at any time and $\boldsymbol{\sigma}_S^0$ is the nonzero stress tensor in the reference configuration (diastole). Moreover, $\boldsymbol{\varepsilon}_s$ is the strain tensor derived from such as $\boldsymbol{\varepsilon}_s = (\nabla \mathbf{u}_s + {}^t\nabla \mathbf{u}_s)/2$, $\mathbf{I}$ is the identity tensor, $E$ is the linearized Young's modulus and $v$ is the Poisson ratio, taken equal to 0.49 (quasi incompressibility). The momentum equation which governs solid dynamics and relates spatial and temporal variations of $\boldsymbol{u}_s$ and $\boldsymbol{\sigma}_s$ may be written as:

$$\rho_s \frac{\partial^2 \boldsymbol{u}_s}{\partial t^2} - \nabla \cdot \boldsymbol{\sigma}_s = 0 \qquad (5)$$

where $\rho_s$ is the wall density. To find an approximate solution to Eq. 5, the solid domain was partitioned in tetrahedral finite elements using Ansys Mechanical (ANSYS Inc. Canonsburg, PA, USA). A mesh refinement analysis was carried out to ensure a mesh-independent solution and to find an optimal compromise between efficiency and accuracy of the results. Two meshes were tested in terms of peak wall stress: a coarse mesh (6077 elements and 2130 node) and a fine mesh (18141 elements and 6184 nodes). Mesh-independency was achieved for an error lower than 5%. The solution $\boldsymbol{u}_s$ was searched in a subspace of finite dimension generated using quadratic shape functions. The average size of tetrahedral finite-elements was 3.5 mm.

Boundary conditions were assigned at every inlet and outlet such as $\boldsymbol{u}_s \cdot \boldsymbol{n}_{out} = 0$ (only radial displacements are allowed).

The luminal wall (interface between the fluid and solid domains) was assigned a Neumann boundary condition coupling the fluid and the solid domain such as:

$$(\boldsymbol{\sigma}_s - \boldsymbol{\sigma}_S^0) \cdot \boldsymbol{n}_s + (\boldsymbol{\sigma}_f - p_{DIAS}\boldsymbol{I}) \cdot \boldsymbol{n}_f = 0 \qquad (6)$$

where $p_{DIAS}$ is the diastolic pressure, equal to 85 mmHg. The introduction of $-p_{DIAS}\boldsymbol{I}$ in the boundary conditions balances $\boldsymbol{\sigma}_S^0$ at diastole, permitting to ensure $\boldsymbol{u}_s \approx \boldsymbol{0}$ at diastole. This condition was required as the reference geometry of the fluid and solid domain were reconstructed at diastole so no displacement should be expected in the diastolic state. From a numerical point of view, the introduction of $-p_{DIAS}\boldsymbol{I}$ also improved the stability of FSI simulations [37]. Indeed, the surrounding tissue had a damping effect on the motion of the aortic wall [38]. After discretizing Eq. 5 in Ansys Mechanical, a Rayleigh damping was introduced such as:

$$[\boldsymbol{C}] = \alpha[\boldsymbol{M}] + \beta[\boldsymbol{K}] \quad (7)$$

where $[\boldsymbol{M}]$ is the mass matrix and $[\boldsymbol{K}]$ the stiffness matrix.

### 2.3 FSI system coupling

The simulations were performed using ANSYS Fluent v17.2 for the fluid and ANSYS Mechanical for the solid. The interaction between the two domains took place at the interface between the solid domain and the fluid domain. At this interface, the fluid pressure was transferred to the solid domain (Eq. 6) and the displacement of the solid walls was transferred to the fluid (Eq. 3). The two systems were connected by the System Coupling component which controls the solver execution of fluid and solid simulations in ANSYS. Fluid and solid domains were solved sequentially starting from the fluid simulations. Time steps were divided into coupling iterations, and for each coupling iterations, ANSYS Fluent passed the loads on the wall interface to ANSYS Mechanical, which in turn transferred back the mesh deformations to ANSYS Fluent [39]. The coupling iterations were repeated until the convergence was reached or a new time step was run. To increase the stability and convergence of the FSI simulations, a relaxation factor for the loads passed between ANSYS Fluent and ANSYS Mechanical, named the under-relaxation factor, was tuned. Finally, a maximum root-mean-square (RMS) residual of 0.01 had to be reached for both fluid and solid domains to ensure the convergence of the solution. A complete simulation took an average of 5 days to be processed on a quad-core Intel® Core™ i5-4590 CPU machine with 16 GB of RAM.

### 2.4 Verification

#### 2.4.1. Sensitivity analysis on convergence speed

A reference simulation (denoted case 1 onwards) was defined with the physiological patient's parameters and a sensitivity analysis was performed on different parameters affecting the convergence speed of the simulations.

The three-element Windkessel model was tuned to reach the desired pressure and flow waveforms measured by 4D MRI. The aortic characteristic impedance (Z) was equal to $6.6\ 10^6$ kg·m$^{-4}$·s$^{-1}$, the peripheral resistance (R) was of $1.6\ 10^9$ kg·m$^{-4}$·s$^{-1}$ and the total arterial compliance (C) was equal to $7.1\ 10^{-9}$ kg$^{-1}$·m$^4$·s$^2$.

Based on previous studies about wall mechanical properties, the physiological linearized Young's modulus of the wall was set to $E$=2 MPa.

The sensitivity to the average mesh size in the fluid domain was first characterized. The error was calculated as the relative difference of the velocity at the distal outlet (DescAo) between the coarser (average element size = 1.5 mm) and the finer mesh (average element size = 1 mm). A relative error smaller than 2% was considered acceptable. Furthermore, on top of the average mesh size, we also investigated the role played by mesh quality indicators such as the minimum orthogonal quality and the maximum aspect ratio. Poor mesh quality was often responsible for divergence of the solution due to negative cell volumes in the fluid domain. The negative cell volumes were elements in which the vertices were inverted because the displacement resulting from the moving mesh velocity $w$ in an iteration was larger than the size of the smaller element.

Mesh convergence was also investigated for the solid domain, where the mesh error was calculated as the relative difference of the maximum principal stress between the coarser (element size 6.5 mm) and the finer mesh (element size 3.5 mm). A relative error smaller than 5% was considered acceptable. On top of the average mesh size, we also tried to optimize mesh quality indicators to avoid highly distorted elements causing divergence of the mechanical solution. Convergence was also improved by calibrating the Rayleigh damping parameters $\alpha$ and $\beta$.

We also varied the numerical parameters of the transient analysis: time step size, maximum number of iterations, maximum number of substeps, under-relaxation factor (factor between 0 and 1 which reduces the increments of variables produced during each iteration). The setting of the under-relaxation factor was especially critical: a high value led to numerical instabilities, whilst too low value significantly slowed down convergence.

### 2.4.2. Verification against 4D MRI datasets

After the sensitivity analysis on convergence speed, the velocity maps and the velocity profiles in the dilated region of the aTAA obtained from the FSI analysis of case 1 were compared to the 4D MRI velocities. Two planes of interest were defined (aTAA$_{middle}$ in the region of the maximum dilatation and aTAA$_{end}$ in the region downstream the aneurysm, Figure 1A) by taking into account the diameter at these planes and the distance from the aorta inlet. The flow eccentricity (Flow$_{eccentricity}$) was defined by the Euclidean distance between the vessel centerline and the velocity center of the forward flow normalized to the lumen diameter, as following:

$$\text{Flow}_{\text{eccentricity}} = \frac{\sqrt{\Sigma_j (C_j - \text{Cvel}_j)^2}}{D} \qquad j=x,y,z \qquad (8)$$

where $C_j$ is the coordinate of the center of the lumen, $\text{Cvel}_j$ is the 'center of velocity' and D is the diameter. The "center of velocity", $\text{Cvel}_j$, was calculated as the weighted barycenter of the cross section, each position being weighted by the velocity value:

$$\text{Cvel}_j = \frac{\Sigma_i r_{i,j} |v_i|}{\Sigma_i |v_i|} \quad i = \text{lumen pixel} \quad j=x,y,z \qquad (9)$$

where r is the radius and v is the velocity.

$\text{Flow}_{\text{eccentricity}}$ equal to 0 indicates that flow is centrally distributed with respect to the vessel centerline. $\text{Flow}_{\text{eccentricity}}$ equal to 1 indicates that the flow is eccentric and impinges against the vessel walls [40]. The $\text{Flow}_{\text{eccentricity}}$ value was calculated at the systolic peak (time = 0.2 s) from the FSI simulations and was verified against the results obtained from the 4D MRI datasets.

### 2.5 Sensitivity to aortic stiffness and peripheral resistance

After the sensitivity analysis on convergence speed and verification against 4D MRI data for the reference simulation (case 1), different other cases were simulated to evaluate the influence of an acute change of peripheral aortic resistance or the impact of aortic stiffness on the aTAA risk of rupture. In this paper we report the 3 following ones:

Case 2 has the same aortic wall stiffness ($E$=2 MPa) but a 10 fold peripheral resistance ($R$=1.6 $10^{10}$).

Case 3 has a 5 fold aortic wall stiffness ($E$=10 MPa) with normal physiological peripheral resistance ($R$=1.6 $10^9$).

Case 4 has a 5 fold aortic wall stiffness ($E$=10 MPa) with 10 fold peripheral resistance ($R$=1.6 $10^{10}$).

The distribution of the blood flow, the distribution of the WSS and the stress distributions in the wall were evaluated for each case at the systolic peak (time 0.2s). The different parameters for each case are summarized in Table 1.

## 3. Results

### 3.1 Sensitivity analysis on computational parameters

The optimal numerical parameters providing a reasonable compromise between the accuracy of the solution and computational cost are summarized in Table 2. The average mesh size in the fluid domain was 1 mm and the average mesh size in the solid domain was 3.5 mm. The following quality mesh indicators were reached: for the fluid domain, a maximum aspect ratio of 34 with an average of 5 was obtained and a minimum orthogonal quality of 0.17 with a mean value of 0.83 was reached (where a value of 0 is worst and a value of 1 is best [31]); for the solid domain, a minimum orthogonal quality of 0.18 with a mean value of 0.71 were reached. It was observed that a mesh quality beyond 0.4 was required in the fluid domain to avoid divergence of the solution due to negative cell volumes.

Optimal parameters for the Rayleigh damping were found to be $\alpha$ = 5650 and $\beta$ = 0.1 [38].

It was found that a time step of 0.001 s gave a convergent solution and smaller time steps did not modify the solution. A number of 20 substeps was imposed in Ansys Mechanical and a maximum number of 200 iterations was set in Fluent. An under-relaxation factor equal to 0.3 was considered and 10 iterations were used, which means that at each iteration, the under-relaxation factor took part of the solution value from previous iteration to dampen solution and cut out steep oscillations, increasing the stability of the calculation. This was considered as a good compromise between numerical instabilities and computational cost.

### 3.2 Verification against 4D MRI datasets

Flow$_{eccentricity}$ at the systolic peak (time = 0.2s) calculated from the CFD studies was verified against the results obtained from the 4D MRI analysis for the reference simulation (case 1) (Table 3). The highest Flow$_{eccentricity}$ was found in the region of the bulge (aTAA$_{middle}$, Figure 1A). There was a fairly good agreement between Flow$_{eccentricity}$ obtained from the CFD simulations and Flow$_{eccentricity}$ obtained in 4D MRI (a difference in percentage of 22% for the aTAA$_{middle}$ and 28%, for the aTAA$_{end}$). Both indicate a deviation of the velocity flow away from the aortic centerline and a jet flow impingement against the aortic wall (Figure 4). Large WSS were also found in this region (Figure 5). Although we compared only 4D MRI datasets to case 1, it was observed that Flow$_{eccentricity}$ remained the same in the other 3 cases, which confirmed that hemodynamics was mostly driven by the geometrical factors (such as the aneurysm bulge, shape, tortuosity and twist).

### 3.3 Sensitivity to aortic stiffness and peripheral resistance

In Table 4, we report the pressure and the peak wall stress (first principal stress) for each case of different aortic stiffness and peripheral resistance. The results are reported at the systolic peak (0.2 s). We also report the peak of the membrane stress, which is defined as the average of the first principal stress across the wall thickness. Note that the stresses are reported as the difference with respect to the diastolic stress.

Figure 4 shows the maps of blood pressure. High blood pressure values were found in the anterior region of the ascending aorta near the greater curvature of the aTAA for all the four cases. The maximum pressure was equal to 18kPa for case 1, 30 kPa for case 2, 20kPa for case 3 and 62kPa for case 4. A significant increase of blood pressure was found in case 4 for the larger peripheral resistance and wall stiffness: 62 kPa = 450 mmHg.

The increase of aortic stiffness had a major impact. It induced a peak wall stress almost 7 times higher (from 105 kPa for the normal peripheral resistance to 702 kPa for the higher peripheral resistance) and a pressure approximately 3 times higher (20 kPa against 62 kPa); whereas between case 1 and case 2 (wall stiffness = 2 MPa) the peak wall stress was increased by a factor 3.5 (from 77 kPa to 260 kPa for the higher peripheral resistance) and the pressure by a factor 1.5 compared to the normal peripheral resistance (18 kPa versus 30 kPa).

In every case, the peak wall stresses were located on either the anterior and posterior regions of the ascending aorta (Figure 6). However, stresses on the posterior side tended to be slightly higher than those on the anterior side, increased by a factor of 1.2 in every case, as it is reported in the Table 5.

Finally, the membrane stress presented the same trend as the peak wall stress. To conclude, our results suggest that patients with a stiffer aTAA may reach very high peak wall stress in case of acute rise of peripheral resistance whereas patients with a more compliant aTAA keep moderate stresses for similar acute rise of peripheral resistance. This shows that the risk of rupture of aTAA is significantly increased with aTAA stiffening.

## 4. Discussion

In this study, we evaluated computationally using FSI analyses the role played by the variation of peripheral resistance and the impact of aortic stiffness onto peak wall stress in aTAA. Our findings suggest that stiffer aTAA may have the most altered distribution of wall stress and an acute change of peripheral vascular resistance could significantly increase the risk of rupture for stiffer aneurysms.

A number of computational studies [2, 41] have already been dedicated to aTAA, most of them aimed at deriving hemodynamic descriptors (i.e., blood pressure, flow patterns and WSS) to identify pathological disturbances leading to vessel dilatation and aneurysm development. Indeed, it is commonly admitted [13, 7] that pathogenesis of aTAA is associated to disturbed blood flows combined with genetic or developmental defects in the proximal aortic tissue, leading to weakness of aortic wall and risk of aneurysm formation.

Although it is confirmed that disturbed aortic flow predisposes the ascending thoracic aorta to aneurysm, wall stress is also widely acknowledged to render the aorta susceptible to the initiation of an aortic dissection. In the last two decades, several studies proposed the peak wall stress as an indicator to predict the risk of rupture of aneurysms. However, most of these studies focused on abdominal aortic aneurysm (AAA) [42, 43]. Among these, Wilson et al. [44] reported a relationship between the aortic wall distensibility and AAA rupture, showing that a decrease in stiffness (increase in distensibility) was related to a shorter time to rupture, independently of other risk factors. However, the physiopathology, genetics and biomechanics of each type of aneurysm are known to be different [17, 45-47]. It was also shown that vSMCs isolated from the thoracic aorta respond differently than vSMCs isolated from the abdominal aorta [47], which may be explained by their different embryonic origins. Given these differences, our results for aTAA do not necessarily extend to AAA. Nevertheless, even if AAAs and aTAAs can arise from different etiologies, their rupture can be modeled similarly with the concept of biomechanical failure [17]. This concept states that rupture or dissection occurs when the peak wall stress in the tissue reaches the maximum stress, or strength of the tissue.

Most studies dedicated to deriving patient-specific wall stress distributions in aTAA used a quasi-static pressure or overpressure loading [14, 48-50]. Trabelsi et al. [14] developed patient-specific finite element models and estimated the wall stress distribution of 5 human aTAAs at systolic pressures and showed that the peak wall stress was located on the inner curvature of the aneurysm. The peak wall stress could reach values over 500 kPa at hot spots of the inner curvature. These results were also confirmed by Mousavi et al. [48], who proposed a layer-specific damage model to computationally predict the risk of tear formation. Alford & Taber [51] had shown earlier that in the aortic arch, like in a torus, compared to the basal circumferential stress of a cylinder of similar diameter, the circumferential stress at the inner curvature increases while the stress at the outer curvature decreases.

Pasta et al. [12] and Khanafer & Berguer [52] were among the few groups to conduct FSI analyses on aTAA. Pasta et al. [12] evaluated hemodynamic predictors and wall stresses in patients with aTAA including both BAV and TAV genotypes, taking into account a bi-layered aorta with material properties obtained from tensile tests for each layer (intima-media and adventitia). They also found peak wall stress located on the inner curvature. They reached stress larger than 3 MPa in the media layer, which was shown to take much larger stresses than the adventitia. Finally, all previous studies, including the current one, share the same location of peak wall stress in aTAAs.

The location of peak wall stress is not systematically the most prone to rupture. Indeed, rupture occurs where the wall stress reaches the strength of the tissue according to a relevant failure criterion. We used the failure criterion of maximum principal stress. Several studies demonstrated evidences of regional differences in the strength of aTAAs as well as non-uniform distribution of tissue thickness [19, 12, 16, 53]. Therefore, it may happen that even if the peak wall stress is located on the inner curvature side, a dissection can initiate on the outer curvature side where the tissue may be weaker due to the effects of disturbed hemodynamics. Moreover, the initiation of dissection is very complex and would require to establish adapted relevant failure criteria [48].

There is no real consensus about the relevant criterion which should be used to predict biomechanically the risk of rupture of aTAA. For instance, Martin et al. [24] defined a new rupture risk criterion (the diameter ratio risk) as the ratio between the current diameter of the aneurysm and the rupture diameter. They showed that an elevated yield diameter ratio risk is significantly associated with increases of the physiological elastic modulus of the artery. Our research group [21] proposed a similar rupture risk criterion, namely the stretch ratio risk, defined as the current tissue stretch (circumferential and axial component, under *in vivo* conditions) and the maximum stretch (at which the tissue ruptures). The rupture risk criterion was obtained with bulge inflation tests on a cohort of 31 patients undergoing elective surgical repair. Moreover, from these tests we derived the tangent elastic modulus of the aTAA tissues and we demonstrated the strong correlation to the stretch ratio risk creation [21]. This result highlighted the relationship between stiffness and rupture susceptibility. Once the aneurysm process begins, the proteolytic activity increases, which leads to remodeling and increasing of wall stiffness. Any acute increase in blood pressure could significantly increase wall stress and make aneurysmal vessels more susceptible to rupture. Indeed, Martin et al. [24] and Hatzaras et al. [25] reported that patients with Marfan syndrome should avoid strenuous activity that increase blood pressure acutely, such as weight lifting, as rupture or dissections in aTAA often occur at a time of severe emotional stress or physical exertion [24]. However, benefit derived from dynamic exercise have been discussed recently. Les et al. [54] investigated hemodynamics under rest and exercise conditions in eight AAA patients. The studies demonstrated that exercise may positively alter the hemodynamic conditions assumed to induce aneurysm growth: the low, oscillatory flow seen at rest, which is hypothesized to be associated with aneurysm growth, was largely eliminated during exercise. This eventually shows that the rupture risk cannot be assessed simply by considering the quasi-static effect of the blood pressure, but should consider exceptional loadings and exercise as well.

This was the motivation to set up the FSI model of the current paper and to refer to the physiological condition of hypertension which manifests with sudden increase of peripheral resistance and induces temporarily significant changes of blood volumes in the aorta [26]. To the best of our knowledge, the influence of an acute change of peripheral aortic resistance or the impact of aortic stiffness on the aTAA risk of rupture had never been assessed computationally before. This is the first biomechanical investigation taking into account the consequences on the risk of aTAA rupture or dissection due to a wall stiffness variation and/or a sudden change of the peripheral vascular resistance. Our findings suggest that a stiffer aneurysm presents higher risks of rupture during situations of pathological condition of hypertension.

## 5. Limitations

Several limitations are still present in this work and should be addressed in the future.

First, the FSI models were obtained from the 4D MRI scan of a single patient. A cohort of patients should be considered to confirm our conclusions. In addition, the patient of the current study had a bicuspid aortic valve; a larger study should include simultaneous BAV and TAV patients, as the wall stress in both groups may show different patterns [12].

Secondly, direct numerical simulations (DNS) would be required to properly resolve turbulences. However, according to previous analysis [55] the time-averaged wall shear stress (TAWSS) distribution between high resolution (HR) and Normal Resolution (NR) simulations are comparable. As the focus of the study is to evaluate the stress in the aortic walls rather than the WSS, we preferred to opt for NR simulations.

Thirdly, an external counter pressure was applied on the outer aortic walls of the segmented geometry, permitting to use a linearized elastic behavior for the aortic wall. An alternative approach would be to derive the unloaded geometry at zero pressure from the segmented diastolic 3D model using the backward-incremental method [14], but this would require to model the hyperelastic behavior of the aortic wall through the whole range of strains spanned between the zero pressure and the systolic pressure.

In addition, tissue thickness and material parameters were assumed uniform for the entire aorta.

The wall thickness was assumed uniform with a value of 1.5 mm according to previous measurements [21]. Experimental studies have demonstrated that the wall thickness changes across the arterial tree [16, 56]. It was shown that neglecting variations of wall thickness can lead to underestimation of the wall stress up to 20% [14, 57]. Some authors reported smaller thickness values on the outer curvature side of the ascending thoracic aorta [19] which may counterbalance the smaller stresses found on that side of the aorta. However, 4D MRI imaging does not allow identifying the aortic wall thickness. Recognizing local material properties is even more challenging and it is a topic of ongoing research. Future studies could combine thickness, directional wall properties variation and wall stress together in wall tensions [58].

Finally, neither axial pulling nor twisting movements applied by the heart on the ascending thoracic aorta was taken into account. According to Mousavi et al. [15], including the heart natural movements in the model can slightly increase the maximum principal stress in the aTAA wall. This should be considered in future studies.

## 6. Conclusions

In this paper, a patient-specific FSI model was employed to analyze the hemodynamics and the biomechanics in case of aTAA. 4D MRI was used to assign boundary conditions and to validate the model. The objective was to evaluate computationally the role played by the variation of peripheral resistance and the impact of aortic stiffness onto peak wall stress in aTAA. Our findings suggest that a stiffer aTAA may have the most altered distribution of wall stress and an acute change of peripheral vascular resistance could significantly increase the risk of rupture for a stiffer aneurysm. This is the first time that the risk of rupture of an aTAA is quantified in case of combined effects of hypertension and aortic stiffness increase. Acute rise of peripheral resistance in hypertension has been reported as a common cause of aneurysm rupture or dissection [59]. Therefore, it can be concluded that a stiffer aneurysm present higher risks of rupture due to hypertension. Future work will extend the study to a cohort of patients including BAV and TAV patients. Moreover, model refinement will consider aortic root motion throughout the cardiac cycle for a more precise assessment of aTAA wall stresses.

## 7. Compliance with ethical standards

Funding: This research was supported by the European Research Council (ERC grant biolochanics, grant number 647067, grant holder: SA).

Conflict of interest: All the authors declare they have no conflict of interest.

Ethical approval: All procedures performed in this study were in accordance with the ethical standards of the 1964 Helsinki declaration and its later amendments. The study was approved by the Institutional Review Board of the University Hospital center of Saint-Étienne (France). After informed consent, a 59-year-old man was enrolled. The patient was scanned on a 3T MRI scanner (Siemens Magnetom Prisma) without contrast agent using a 4D flow phase contrast protocol and sequence.

## 8. Acknowledgments

We thank Dr Morbiducci and Dr Gallo from Polytechnic of Turin (Italy) who provided insight and expertise that greatly assisted the model development. We are also grateful to Ansys, Inc. for providing Ansys-Fluent (ANSYS® Academic Research, Release 17.2). This research was supported by the European Research Council (ERC grant biolochanics, grant number 647067).

**List of tables:**

**Table 1.** Parameters employed for each case analyzed in order to evaluate the influence of an acute change of the peripheral aortic resistance and/or the changes in the vessel wall stiffness on the aTAA risk of rupture.

**Table 2.** Numerical parameters recommended to provide a reasonable compromise between the accuracy of the solution and computational cost.

**Table 3.** $Flow_{eccentricity}$ calculated from the CFD studies against the 4D MRI results for the reference simulation (case 1), at the systolic peak.

**Table 4.** Pressure, peak wall stress and peak membrane stress results obtained at the systolic peak (time= 0.2s). The worst case scenario for pressure and peak wall stress was found for case 4, in which the highest wall stiffness and peripheral resistance were used. Finally, peak membrane stress presented the same trend as the peak wall stress.

**Table 5.** Peak wall stress at the anterior and posterior regions of the ascending aorta. Stresses on the posterior side tended to be slightly higher compared to those on the anterior side.

**Table 1**

|  | E (MPa) | R (kg.s.m$^{-4}$) |
|---|---|---|
| Case 1 | 2 | 1.6 10$^9$ |
| Case 2 | 2 | 1.6 10$^{10}$ |
| Case 3 | 10 | 1.6 10$^9$ |
| Case 4 | 10 | 1.6 10$^{10}$ |

**Table 2**

| Ansys Fluent | Mesh size = 1 mm |
| | Mesh quality = 0.4 |
| Ansys Mechanical | Mesh size = 3.5 mm |
| | Rayleigh Damping: $\alpha$ = 5650  $\beta$ = 0.1 |
| System Coupling | Time step size = 0.001 s |
| | Under relaxation factors = 0.3 |

**Table 3**

| aTAA$_{middle}$ | | aTAA$_{end}$ | |
|---|---|---|---|
| 4D MRI | CFD | 4D MRI | CFD |
| 0.48 | 0.37 | 0.46 | 0.33 |

**Table 4**

| | Case 1 | Case 2 | Case 3 | Case 4 |
|---|---|---|---|---|
| Pressure (kPa) | 18 | 30 | 20 | 62 |
| Peak wall Stress (kPa) | 77 | 260 | 105 | 702 |
| Peak Membrane Stress (kPa) | 46 | 160 | 64 | 440 |

**Table 5**

| | Case 1 | Case 2 | Case 3 | Case 4 |
|---|---|---|---|---|
| Anterior side peak wall stress (kPa) | 61 | 209 | 84 | 565 |
| Posterior side peak wall stress (kPa) | 77 | 260 | 105 | 702 |

**List of figures:**

**Figure 1.** Boundary conditions imposed on the fluid domain. (A) The patient-specific map of velocity profile was obtained from the 4D flow MRI and was used as inflow boundary condition at the aorta inlet (AAo$_{inlet}$). The flow profile in the region of the maximum dilatation (aTAA$_{middle}$) and in the region downstream the bulge (aTAA$_{end}$) are also shown. At the descending aorta (DescAo), a three-elements Windkessel model was considered. (B) The patient specific flow rate was obtained from the MRI analysis and was assigned as outlet boundary conditions to the three apico-aortic branches (BCA, LCC, LSUB). (C) The AAo$_{inlet}$ flow rate waveform resulted from the velocity interpolation along the cardiac cycle.

**Figure 2.** Boundary conditions on the solid domain. (A) Diastolic pressure ($p_{DIAS}$) was applied on the external wall in order to obtain no displacement in the diastolic state. A Rayleigh damping was used. (B) Boundary conditions applied at every inlet and outlets of the model.

Figure 3. (A) CFD results of the velocity contour calculated during the acceleration (t=0.1s) and (B) at the systolic peak (t=0.2s) and verified against the results obtained from the 4D MRI analysis at t=0.1s (C) and at t=0.2s (D).

**Figure 4.** Pressure distribution in Case 1 (A), Case 2 (B), Case 3 (C) and Case 4 (D). For all the cases, high pressure was found in the anterior region of the ascending aorta, near the great curvature of the aneurysm wall. Case 2 (B) and Case 4 (D) showed elevated blood pressure due to high peripheral resistance and wall stiffness.

**Figure 5.** Streamlines of velocity simulated for Case 1 (A), Case 2 (B), Case 3 (C) and Case 4 (D). A jet flow impingement on the anterior region of the ascending aorta was found for all cases.

**Figure 6.** WSS simulated for Case 1 (A), Case 2 (B), Case 3 (C) and Case 4 (D). For all the cases, the peak of WSS was found in the anterior region of the bulge where the jet flow impingement against the aortic wall occurred.

**Figure 7.** Wall stress distribution for Case 1 (A), Case 2 (B), Case 3 (C) and Case 4 (D). For all the cases, the peak of wall stress was located on either the anterior and posterior side of the bulge. Case 4 (D) showed the highest wall stress value due to the highest wall stiffness and peripheral resistance.

**Figure 1**

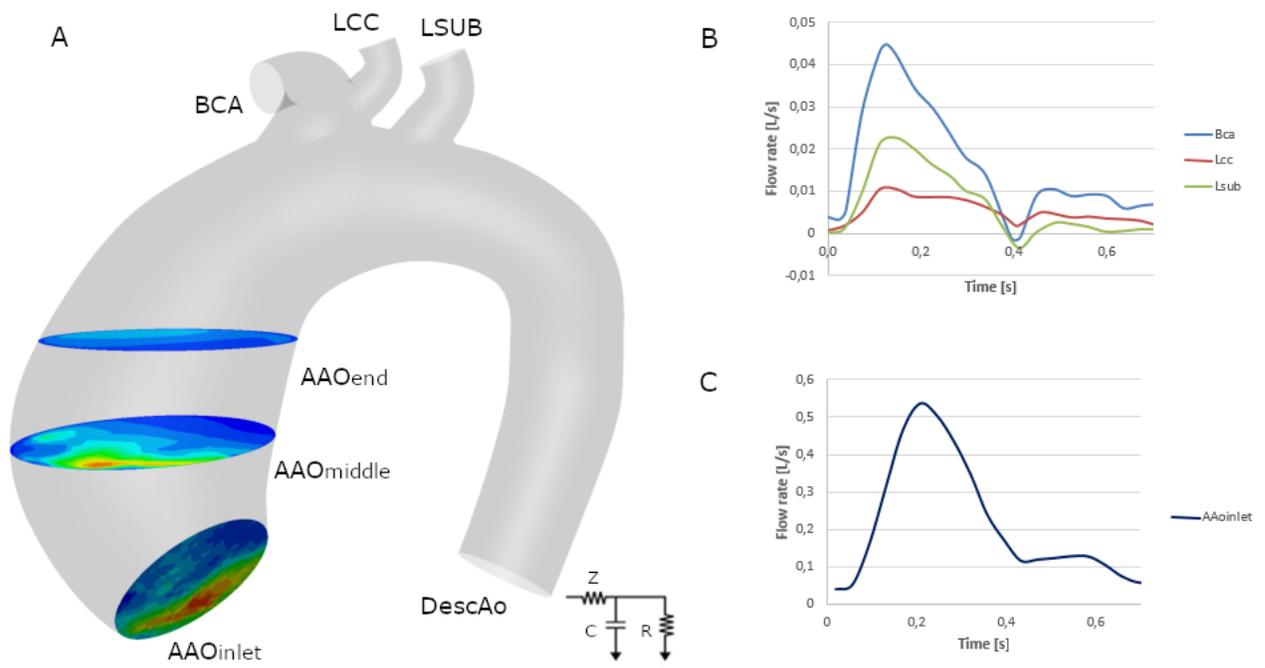

**Figure 2**

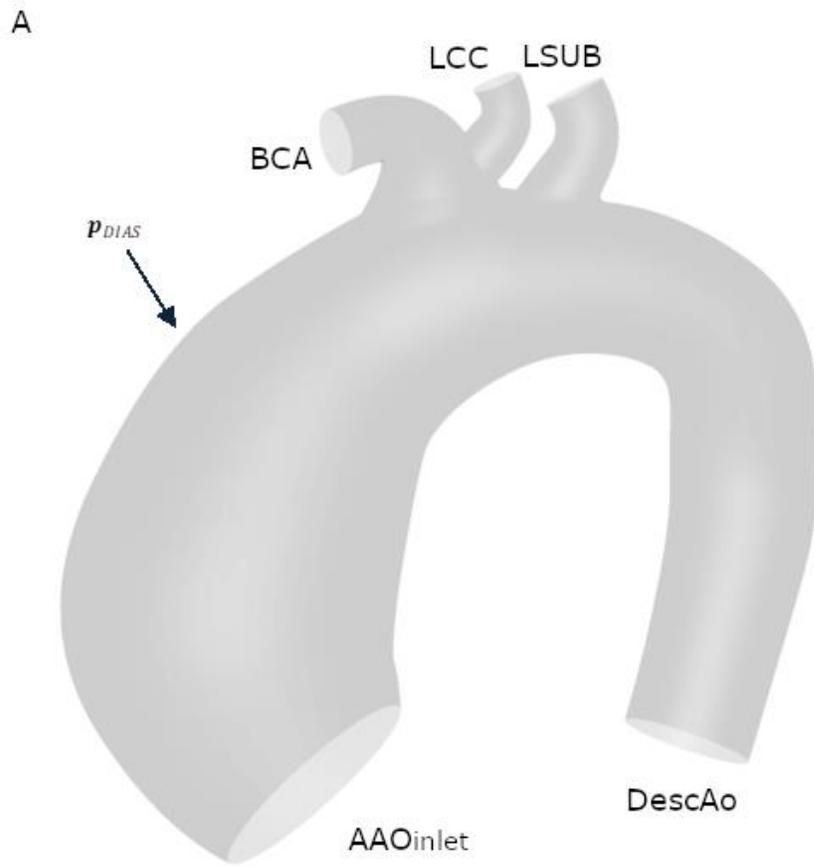
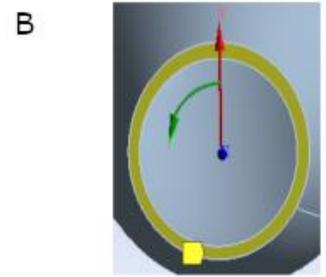

**Figure 3**

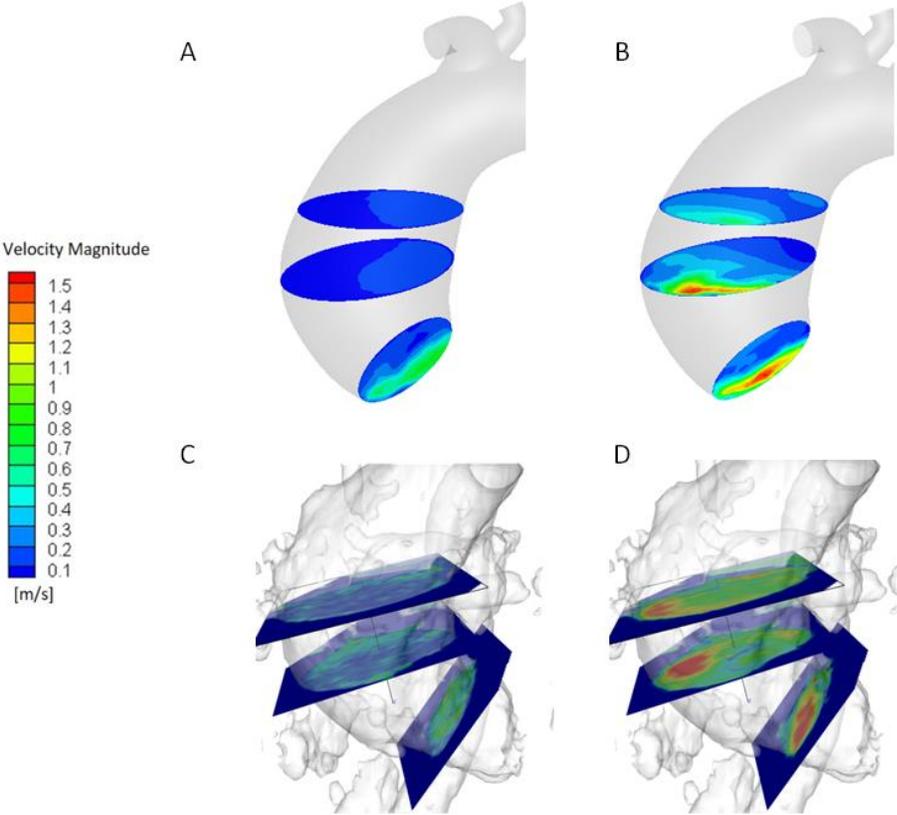

**Figure 4**

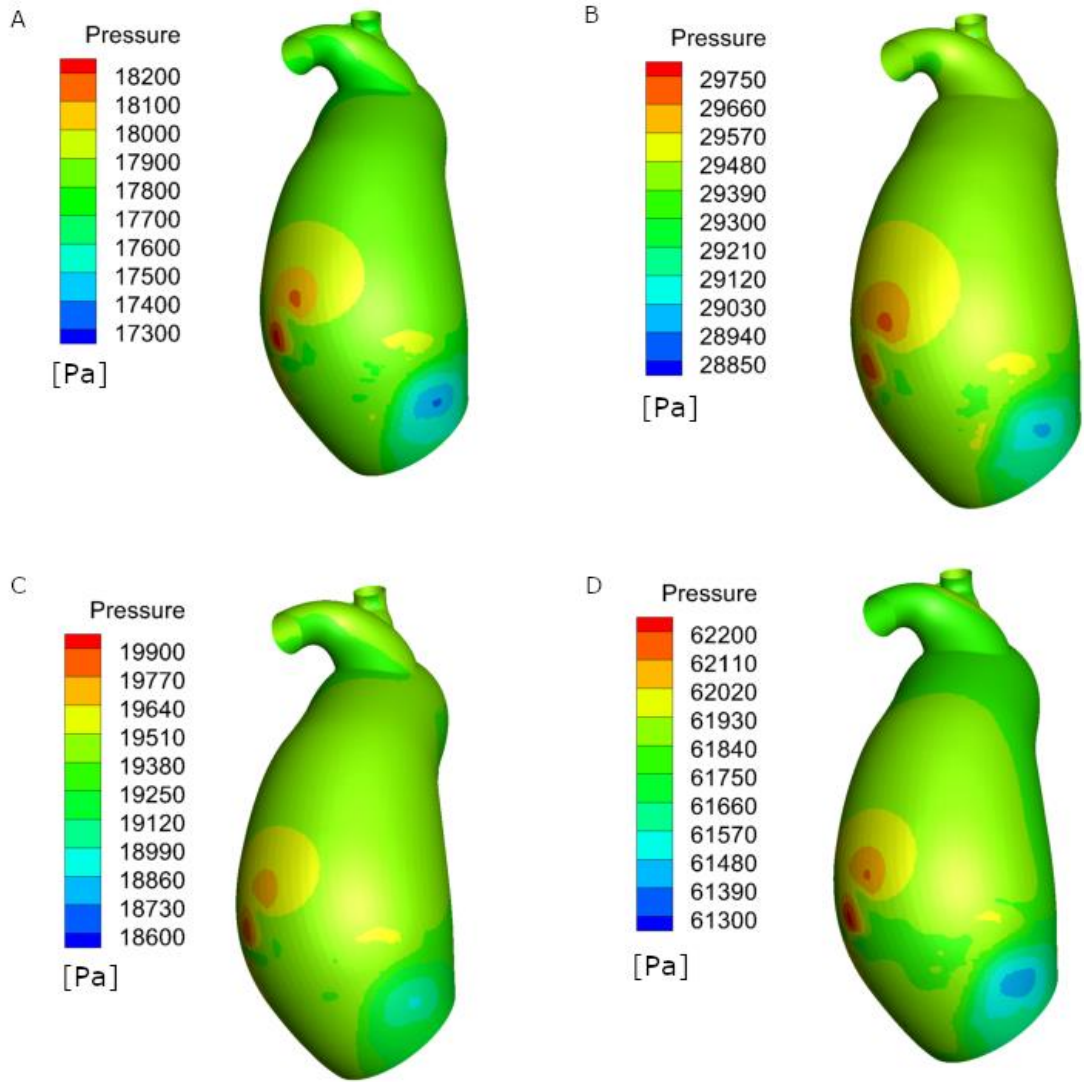

**Figure 5**

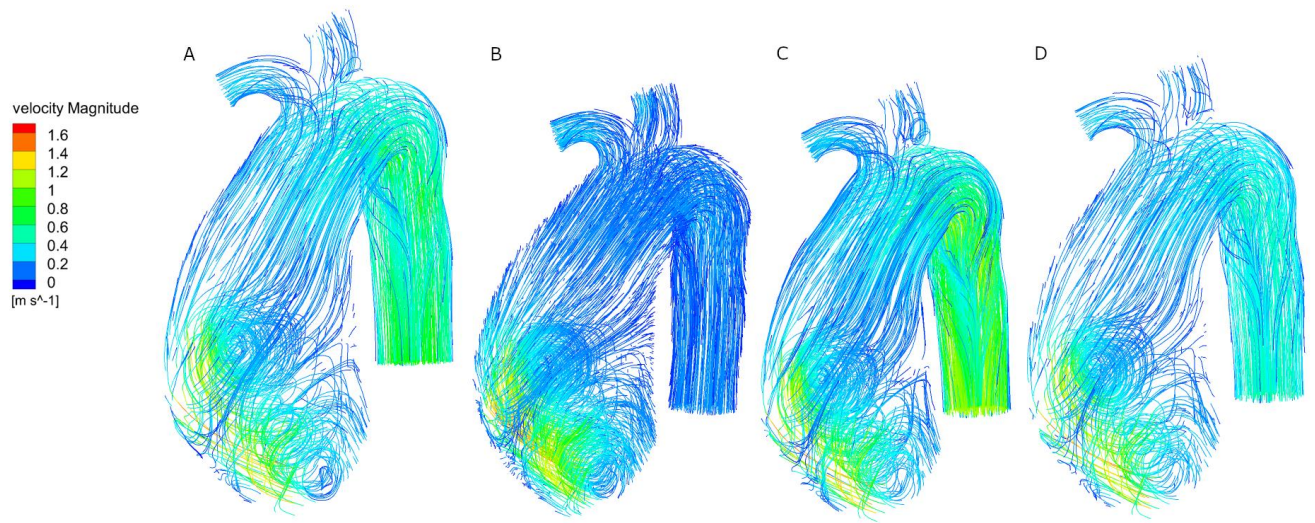

**Figure 6**

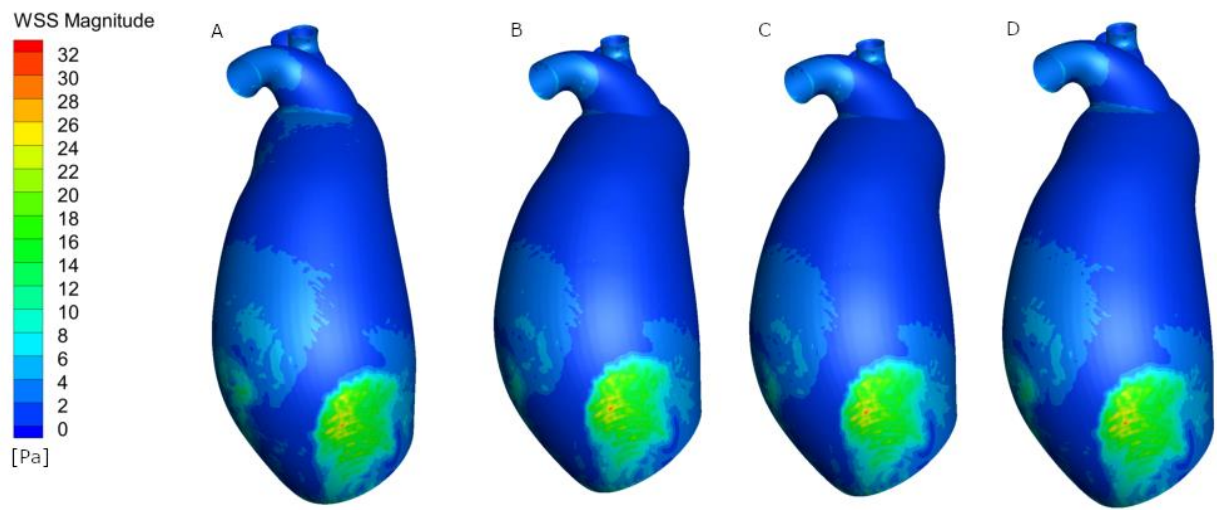

**Figure 7**

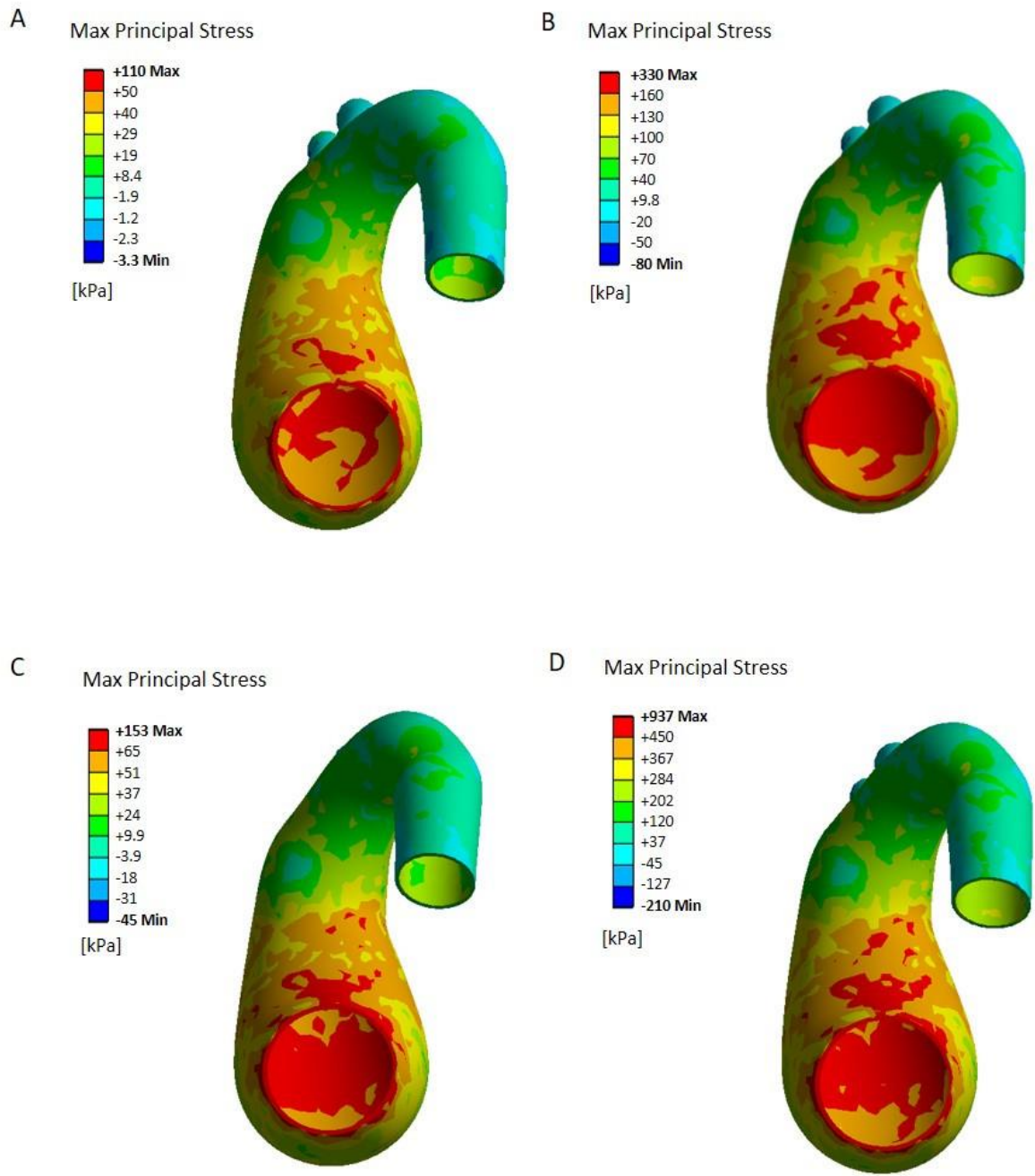